\documentclass{elsart}
\usepackage{amssymb}
\usepackage{graphicx}

\begin{document}
\begin{frontmatter}

\title{Multi-quasiparticle $\gamma$-band structure in neutron-deficient
Ce and Nd isotopes}

\author{ J.A. Sheikh$^{1,2,3}$, G.H. Bhat$^{3}$, R. Palit$^{4}$,
Z. Naik$^{4}$, Y. Sun$^{2,5,6}$ }

\address{$^1$Physics Division, Oak Ridge National Laboratory, P.O. Box 2008,
Oak Ridge, Tennessee 37831, USA \\
$^2$Department of Physics and Astronomy, University of Tennessee,
Knoxville, Tennessee 37996, USA \\
$^3$Department of Physics, University of Kashmir,
Srinagar, 190 006, India \\
$^4$Tata Institute of Fundamental Research, Colaba, Mumbai, 400 005, India\\
$^5$Department of Physics, Shanghai Jiao Tong University, Shanghai
200240, P. R. China\\
$^6$Institute of Modern Physics, Chinese Academy of Sciences,
Lanzhou 730000, P. R. China }

%
%
%


\begin{abstract}

The newly developed multi-quasiparticle triaxial projected shell-model
approach is employed to study the high-spin band structures in
neutron-deficient even-even Ce- and Nd-isotopes.  It is observed
that $\gamma$-bands are built on each intrinsic configuration of the
triaxial mean-field deformation. Due to the fact that a triaxial
configuration is a superposition of several $K$-states, the
projection from these states results in several low-lying bands
originating from the same intrinsic configuration. This generalizes
the well-known concept of the surface $\gamma$-oscillation in
deformed nuclei based on the ground-state to $\gamma$-bands built on
multi-quasiparticle configurations. This new feature provides an
alternative explanation on the observation of two
$I=10$ aligning states in $^{134}$Ce and both exhibiting a neutron character.

\end{abstract}

\end{frontmatter}

\section{Introduction}

One of the main challenges in nuclear structure is to understand the
interplay between single-particle, vibrational and rotational
degrees of excitation modes. Mass $A=120-130$ is one of the regions
where the modes associated with all three degrees of freedom
have been studied quite extensively. High-spin band structure in
this mass region has been experimentally investigated through heavy-ion-induced
reactions (see, for example, refs.
\cite{Pa05,Mason05,Pe96,la05,sa08}). Along with these studies, many
low-lying non-yrast states have also been investigated in the
$\beta$-decay spectroscopic methods \cite{Ga00} and Coulomb
excitations of stable as well as radioactive ions \cite
{Mu06,Ju08,Sa08}. Nuclei in this mass region depict a variety of
band structures which are built on different excitation modes.
Proton and neutron mass distributions develop opposite types of
quadrupole deformations in these nuclei due to the presence of the
proton Fermi surface near the lower end of the $h_{11/2}$ orbitals
and that of neutron near the middle of the $h_{11/2}$ orbitals.
This results in a $\gamma$-instability in these nuclei, and the
$\gamma$-soft behavior is manifested in the observed low-lying
quasi-$\gamma$ bands. This region thus provides an excellent
opportunity to study the effect of rotation and quasiparticle
excitation on the top of $\gamma$-vibrations.

As the aligning neutrons and protons in these rotating systems are
occupying the same intruder orbit $1h_{11/2}$, the neutron and
proton quasiparticle (qp) alignment processes compete in this mass
region. In the majority of the nuclei in this mass region, the proton
alignment occurs earlier than the neutron alignment, which is well
described in the Woods-Saxon TRS framework \cite{Rj88,Wr95}.
However, the neutron alignment is observed to be delayed in many
cases, and the Woods-Saxon TRS analysis always under-predicts this
rotational alignment. There have been other intriguing observations
in this mass region. One problem is in the g-factor measurement in
$^{134}$Ce for the lowest two $I=10$ states, which lie very close to
each other. The issue is that both of these states are shown to have
a neutron character \cite{Ze82} and since these $I=10$ states are
considered to be the bandheads of 2-qp bands, two 2-quasineutron
band structures are observed in $^{134}$Ce. This is quite surprising
as normally lowest-lying two 2-qp states should belong to
2-quasineutron and 2-quasiproton states.

Thus, with the collective and quasiparticle excitations coexisting
in the complex low-lying spectrum in these $\gamma$-soft nuclei, it
is desirable to have a microscopic method that can handle all these
degrees of freedom self-consistently. The early version \cite{SH99}
of the triaxial projected shell model (TPSM) adopted the
triaxially deformed qp vacuum configuration and performed
three-dimensional angular momentum projection. It was shown
\cite{YS02} that one can use this simple configuration to produce
collective multi-phonon $\gamma$-vibrational bands at low spin
states. In this earlier version, mixing with quasiparticle
excitations was neglected and, therefore, it cannot describe
excitation of quasiparticles in a triaxially deformed mean field.
This limitation has been relaxed in our recent development
\cite{Ja08} and the TPSM quasiparticle space for even-even systems
has been considerably extended by inclusion of many 2- and 4-qp
configurations. In a parallel work, a similar extension was done for
odd-odd nuclei \cite{Gao06}. These new developments provide a
suitable shell-model framework to investigate, microscopically, the
current topical issues in nuclear structure that are related to
triaxiality, such as the problem of $\gamma$-bands built on
multi-quasiparticle configurations, discussed in the present work.
They are also applicable to the problems related to the wobbling
motion and to the so-called chiral doublet band structures, which
will be the focus of our future investigations.

The purpose of the present work is to demonstrate that the
traditional picture of $\gamma$-vibration in deformed nuclei, based
on the ground-state configuration, can be generalized to the case of
multi-qp configurations. In the TPSM description, projected states
with $K=0$, 2, and 4 from the 0-qp configuration correspond to
ground, $\gamma$-, and $2\gamma$- bands \cite{YS02}. Similar to
these $\gamma$- and $2\gamma$- bands built on the ground state, new
multi-phonon $\gamma$-bands are based on multi-qp states. For the
multi-qp bands, projection with different $K$-values will correspond
to qp-excited $\gamma$-bands. In particular, we shall show that the
projected $K=1$ and 3 configurations originating from the same
intrinsic neutron 2-qp configuration correspond to the two $I=10$
states observed in $^{134}$Ce. This provides a plausible explanation
to the early observation of the $I=10$ states in this nucleus with
similar intrinsic structure.

The paper is organized as follows: In Section 2, we outline some
basic elements of the TPSM approach. For more details about TPSM, we
refer the reader to ref. \cite{Ja08} and references cited therein. In
Section 3, calculations and discussions on neutron-deficient Ce and
Nd isotopes are presented. Finally in Section 4, we summarize the
present work.

\section{Outline of the Theory}

The extended TPSM qp basis \cite{Ja08} consists of
(angular-momentum) projected qp vacuum (0-qp state), two-proton
($2p$), two-neutron ($2n$), and 4-qp states, i.e.,
\begin{equation}
\{ \hat P^I_{MK}\left|\Phi\right>, \hat P^I_{MK}~a^\dagger_{p_1}
a^\dagger_{p_2} \left|\Phi\right>, \hat P^I_{MK}~a^\dagger_{n_1}
a^\dagger_{n_2} \left|\Phi\right>, \hat P^I_{MK}~a^\dagger_{p_1}
a^\dagger_{p_2} a^\dagger_{n_1} a^\dagger_{n_2} \left|\Phi\right>
\}, \label{basis}
\end{equation}
where the three-dimensional angular-momentum-projection operator is
\begin{equation}
\hat P^I_{MK} = {2I+1 \over 8\pi^2} \int d\Omega\,
D^{I}_{MK}(\Omega)\, \hat R(\Omega), \label{PD}
\end{equation}
with the rotation operator $\hat R(\Omega) = e^{-\imath \alpha \hat
J_z} e^{-\imath \beta \hat J_y} e^{-\imath \gamma \hat J_z}$.
$\left|\Phi\right>$ in (\ref{basis}) is the triaxially-deformed qp
vacuum state. The qp basis chosen above is adequate to describe high-spin
states up to $I\sim 20\hbar$ for nuclei considered in this
work. In the present analysis we shall, therefore, restrict our
discussion to this spin regime.

It is important to note that for the case of axial symmetry, the qp
vacuum state has $K=0$ \cite{KY95}, whereas in the present case of
triaxial deformation, the vacuum state is a superposition of all
possible $K$-values. Rotational bands with the triaxial basis
states, Eq. (\ref{basis}), are obtained by specifying different
values for the $K$-quantum number in the angular-momentum projector
in Eq. (\ref{PD}). The allowed values of the $K$-quantum number for
a given intrinsic state are obtained through the following symmetry
consideration. For $\hat S = e^{-\imath \pi \hat J_z}$, we have
\begin{equation}
\hat P^I_{MK}\left|\Phi\right> = \hat P^I_{MK} \hat S^{\dagger} \hat S
\left|\Phi\right> = e^{\imath \pi (K-\kappa)}
\hat P^I_{MK}\left|\Phi\right>,
\end{equation}
where $\hat S\left|\Phi\right> = e^{-\imath \pi
\kappa}\left|\Phi\right>$. For the self-conjugate vacuum or 0-qp
state, $\kappa=0$ and, therefore, it follows from the above equation
that only $K=$ even values are permitted for this state. For 2-qp
states, $a^\dagger a^\dagger \left|\Phi\right>$, the possible values
for $K$-quantum number are both even and odd, depending on the
structure of the qp state. For example, for a 2-qp state formed from
the combination of the normal and the time-reversed states $\kappa =
0$, only $K$ = even values are permitted. For the combination of the
two normal states, $\kappa=1$ and only $K=$ odd states are permitted.

As in the earlier projected shell-model calculations, we use the
pairing plus quadrupole-quadrupole Hamiltonian \cite{KY95}, with a
quadrupole-pairing term also included:
\begin{equation}
\hat H = \hat H_0 - {1 \over 2} \chi \sum_\mu \hat Q^\dagger_\mu
\hat Q^{}_\mu - G_M \hat P^\dagger \hat P - G_Q \sum_\mu \hat
P^\dagger_\mu\hat P^{}_\mu.
\label{hamham}
\end{equation}
It has been shown by Dufour and Zuker \cite{Dufour96} that these
interaction terms simulate the essence of the important correlations
in nuclei and even the realistic force has to contain, at least,
these basic components implicitly in order to work successfully in
the structure calculations. Some large-scale spherical shell-model
calculations \cite{HK99} also adopt this type of interaction.

The triaxially deformed single-particle basis is obtained from the
Nilsson model \cite{Ni69}. The corresponding triaxial Nilsson
mean-field Hamiltonian is given by
\begin{equation}
\hat H_N = \hat H_0 - {2 \over 3}\hbar\omega\left\{\epsilon\hat Q_0
+\epsilon'{{\hat Q_{+2}+\hat Q_{-2}}\over\sqrt{2}}\right\},
\label{nilsson}
\end{equation}
in which $\epsilon$ and $\epsilon'$ specify the axial and triaxial
deformation, respectively. $\epsilon$ and $\epsilon'$ are related to
the conventional triaxiality parameter by $\gamma =$
tan$^{-1}(\epsilon'/\epsilon)$. In (\ref{nilsson}), $\hat H_0$ is
the spherical single-particle Hamiltonian, which contains a proper
spin-orbit force \cite{Ni69}. The interaction strengths in
(\ref{hamham}) are taken as follows: the $QQ$-force strength $\chi$
is adjusted such that the physical quadrupole deformation $\epsilon$
is obtained as a result of the self-consistent mean-field HFB
calculation \cite{KY95}. The monopole pairing strength $G_M$ is of
the standard form
\begin{equation}
G_M = {{G_1 - G_2{{N-Z}\over A}}\over A} ~{\rm for~neutrons,}~~~~
G_M = {G_1 \over A} ~{\rm for~protons.} \label{pairing}
\end{equation}
In the present calculation, we take $G_1=20.82$ and $G_2=13.58$,
which approximately reproduce the observed odd-even mass difference
in the mass region. This choice of $G_M$ is appropriate for the
single-particle space employed in the model, where three major
shells are used for each type of nucleons ($N=3,4,5$ for both
neutrons and protons). The quadrupole pairing strength $G_Q$ is
assumed to be proportional to $G_M$, and the proportionality
constant being fixed as 0.18.

\section{Results and Discussion}

\begin{table}[t]
\caption{Axial and triaxial quadrupole deformation parameters
$\epsilon$ and $\epsilon^\prime$ employed in the TPSM calculation
for $^{128-134}$Ce and $^{132-138}$Nd isotopes. The corresponding
conventional triaxiality parameter $\gamma$ (in degree) is also
given.}
\begin{tabular*}{105mm}{@{\extracolsep{\fill}}cccc|cccc} \hline\hline
Ce nuclei & $\epsilon$ & $\epsilon'$ & $\gamma$ & Nd nuclei &
$\epsilon$ & $\epsilon'$ & $\gamma$ \\
\hline $^{128}$Ce & 0.250 & 0.120 & 26 & $^{132}$Nd & 0.267 & 0.120 & 24 \\
\hline $^{130}$Ce & 0.225 & 0.120 & 28 & $^{134}$Nd & 0.200 & 0.120 & 31 \\
\hline $^{132}$Ce & 0.183 & 0.100 & 29 & $^{136}$Nd & 0.158 & 0.110 & 35 \\
\hline $^{134}$Ce & 0.150 & 0.100 & 34 & $^{138}$Nd & 0.170 & 0.110 & 33 \\
\hline\hline
\end{tabular*}
\end{table}

\begin{figure}[htb]
\includegraphics[totalheight=8cm]{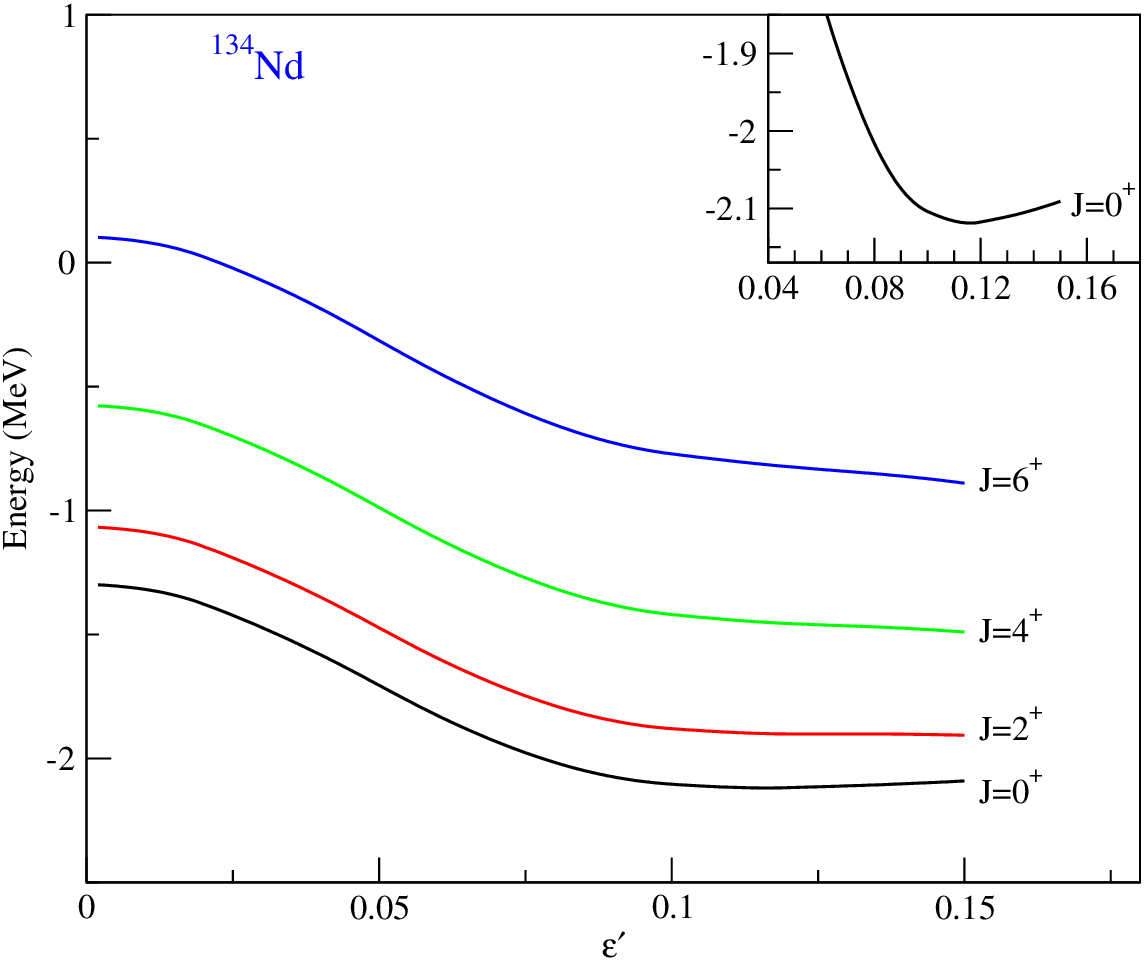}
\caption{(Color online) Calculated projected energy-surfaces for the
low spin states in $^{134}$Nd. } \label{fig1}
\end{figure}

TPSM calculations have been performed for four even-even isotopic chains
of $^{128-134}$Ce and
$^{132-138}$Nd nuclei. 
As already mentioned in the introduction, the main emphasis of the
present work is to perform a detailed investigation of the yrast-
and $\gamma$-bands beyond the first band crossing. The above-mentioned
nuclei were chosen as they have well-developed
$\gamma$-bands observed experimentally and for some of them up to
high spins. The input parameters in the calculation are the
deformation parameters $\epsilon$ and $\epsilon'$, which are given
in Table 1. The corresponding conventional triaxiality parameter
$\gamma$ (in degree) is also given. The quadrupole deformation
$\epsilon$ in the table are those of M\"oller and Nix \cite{Mn95}
(with exception for $^{138}$Nd, see discussion below) and the
triaxiality parameter $\epsilon'$ have been calculated from the
projected potential energy surface. In Fig. 1, we provide one
example from such calculations, in which projected energies as
functions of $\epsilon'$ for a fixed $\epsilon$ ($\epsilon=0.2$ in
the $^{134}$Nd case) are plotted for the low spin states. In this
figure, the $\gamma$-soft nature of the nucleus is clearly seen
since the curves are all flat with triaxiality, specially for
$\epsilon' > 0.08$. As shown in the inset in Fig. 1, a minimum in
the projected ground-state energy is present at $\epsilon'\approx
0.12$. Thus, projected ground-state energies have been calculated
for a range of $\epsilon'$ values, and the value which gives rise to
the minimum energy is used in all further TPSM calculations. We
would like to mention that in one of the studied nuclei, $^{138}$Nd,
ref. \cite{Mn95} predicted a negative value of $\epsilon$
corresponding to an oblate deformation in its ground state, which is
in disagreement with all other neighboring nuclei. However, the TRS
calculation \cite{Liu08} for this nucleus indicates a large
triaxiality of $\gamma\approx 30^o$. In our calculation, we
therefore take a positive $\epsilon$ with a sizable $\epsilon'$ to
construct the basis.

\begin{figure*}[htb]
\includegraphics[totalheight=12.0cm]{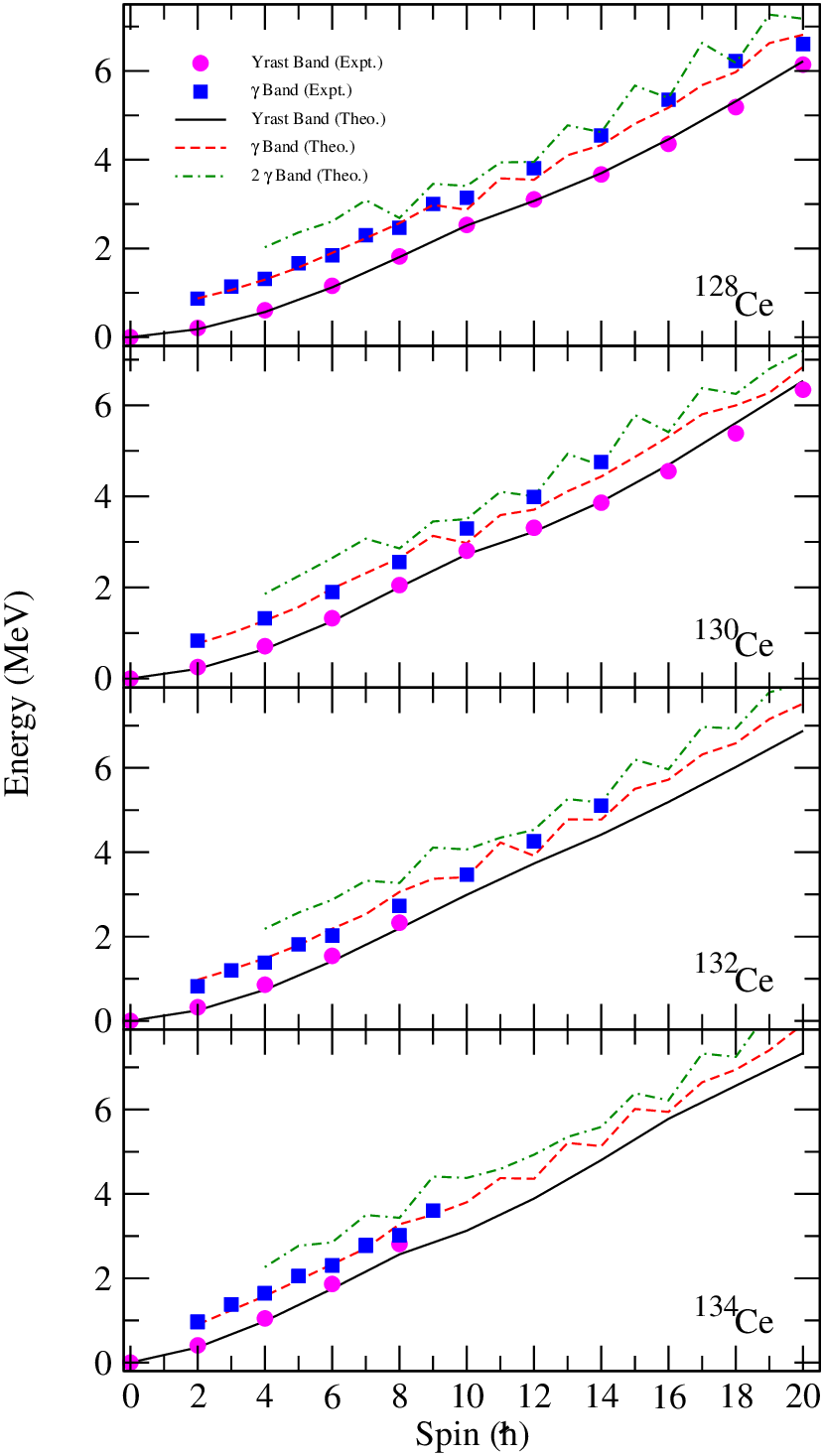}
\includegraphics[totalheight=12.0cm]{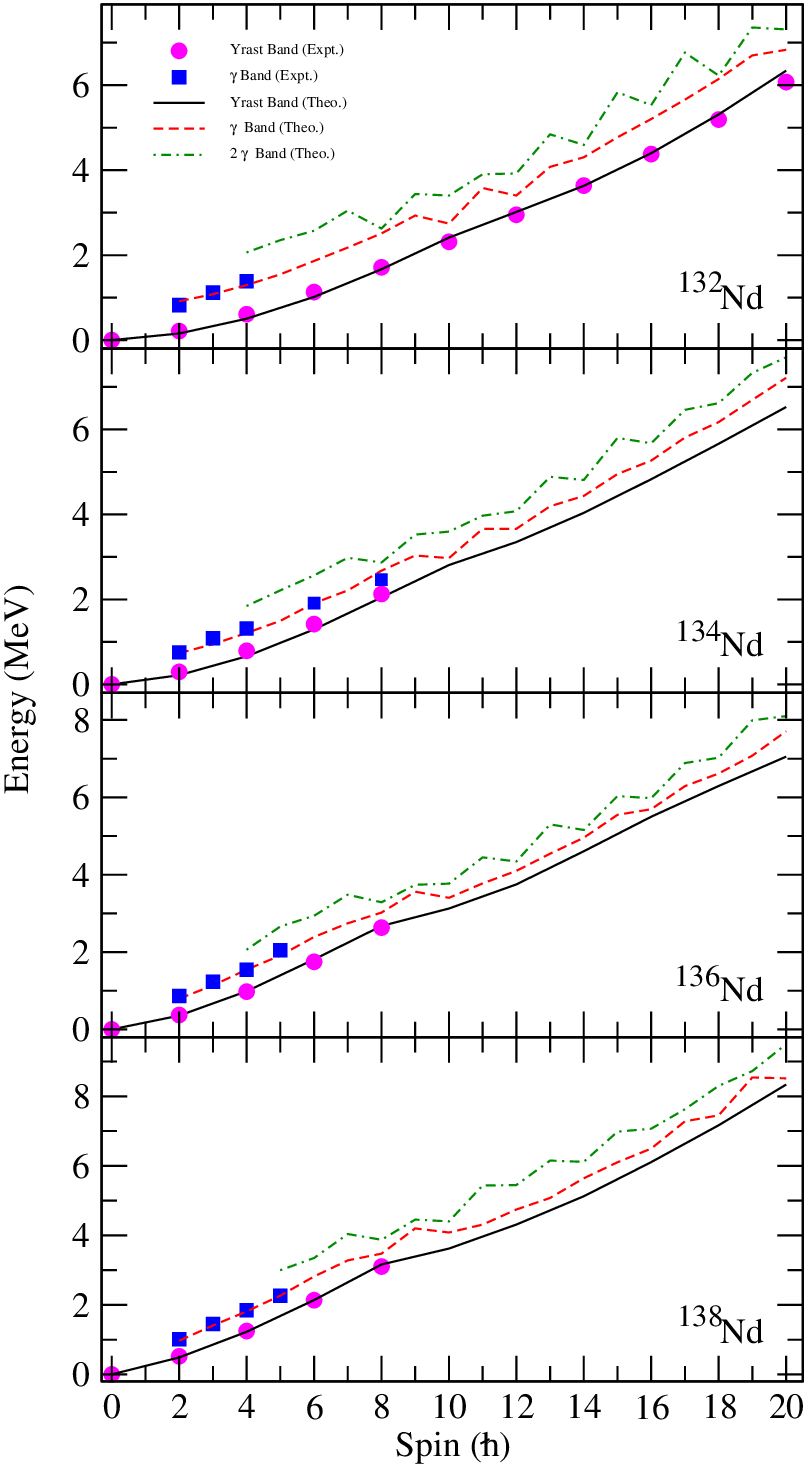}
\caption{(Color online) Comparison of experimental and
calculated band structures for $^{128-134}$Ce and $^{132-138}$Nd.
Data are taken from ref. \cite{Ep00} ($^{128}$Ce), \cite{Ce130}
($^{130}$Ce), \cite{Pa05} ($^{132}$Ce), \cite{Ce134} ($^{134}$Ce),
\cite{Nd132} ($^{132}$Nd), \cite{Cd96} ($^{134}$Nd), \cite{Od02}
($^{136}$Nd), and \cite{Es02} ($^{138}$Nd).} \label{fig2}
\end{figure*}

The calculations are performed in two stages. In the first step, the
projected states are obtained from the triaxially deformed qp states
by applying the three-dimensional angular-momentum projection
method. The projection is carried out for configurations constructed
from various intrinsic states close to the Fermi surface. We have
performed the angular-momentum projection for all the qp
configurations, which are built by considering the single-particle
states that are within 3 MeV from the Fermi surfaces. In the second
stage, the shell-model Hamiltonian, Eq. (\ref{hamham}), is
diagonalized with the projected states as the basis configurations.

The lowest three bands obtained after diagonalization for each
angular momentum are shown in Fig. 2 for all the studied Ce and Nd
isotopes. These bands have the main component from the 0-qp state
and, therefore, are collective bands in the low-spin regime.
Theoretical $K=0$ and $K=2$ bands are respectively compared to the
data of the yrast and $\gamma$-vibrational (one-phonon) bands, and
the $K=4$ bands are our prediction for possible 2$\gamma$
(two-phonon) bands. It can be seen that, overall, an excellent
reproduction for the known data has been achieved. In particular,
the experimental $\gamma$-band in $^{128}$Ce is well reproduced up
to the highest spin state studied in this work. The calculated
$\gamma$-band in $^{136,138}$Nd is predicted to lie very close in
energy to the yrast band, in agreement with known data, and the two
bands become nearly degenerate at high spins.

It is clearly noted in Fig. 2 that yrast bands for all studied
isotopes have two slopes, corresponding to crossings of bands with
two distinct configurations. The change in slope occurs at spin
$I=8$ or 10. This feature has been well described by the
calculation, for the isotopes $^{128,130}$Ce and $^{132}$Nd where
high-spin data exist. Furthermore, we observe that for the rest of
nuclei, i.e., $^{132,134}$Ce and $^{134,136,138}$Nd, the current
ground-band data are available only before the predicted onset of the
band-crossing at $I=10$. The first and second excited bands at
low-spins are predominantly composed of the collective $\gamma$ and
2$\gamma$ structures, but the high-spin states of these bands have
considerable mixing with the 2-qp states. The calculation also
predicts some energy staggerings in the $\gamma$- and 2$\gamma$-
bands. The staggering divides a rotational band with $\Delta I=1$
into two branches with $\Delta I=2$. In the isotopes
$^{128,130,132}$Ce where high-spin data of the $\gamma$-band are
known, we see that the experimental bands actually belong to one of
the branches of the staggering $\gamma$-bands, namely the
energetically favored one with even spins. The present calculation
further predicts some irregularities in the staggering of $\gamma$-
and 2$\gamma$- bands due to band mixing (e.g. staggering appears in
a certain spin range but diminishes at high spins). We hope that our
results can serve as a guidance for future experiments to identify
$\gamma$- and 2$\gamma$- bands in this mass region.

To extract structure information from the calculation, it is useful
to discuss the energies in terms of band diagrams \cite{KY95}. A
band diagram is an ensemble of projected band energies for intrinsic
configurations, i.e., the diagonal matrix elements before band
mixing. It usually shows crossings of various bands where some
prominent phenomena may take place, and therefore plays a central
role in the interpretation of numerical results. One must keep track
of the configurations of each band when plotting a band diagram. As
already mentioned, with the triaxial basis, the intrinsic states do
not have a well-defined $K$ quantum number. Each triaxial
configuration in (\ref{basis}) is a composition of several $K$
values, and bands in band diagrams are obtained by assigning a given
$K$ value in the projection operator. As in ref. \cite{Ja08}, we
denote a $K$ state of an $i$ configuration as $(K, i)$, with $i = 0,
2n, 2p$, and 4. For example, $K = 0$ state of the 0-qp configuration
is marked as (0, 0) and $K = 1$ of the $2n$-qp configuration as $(1,
2n)$.

\begin{figure*}[htb]
\includegraphics[totalheight=12cm]{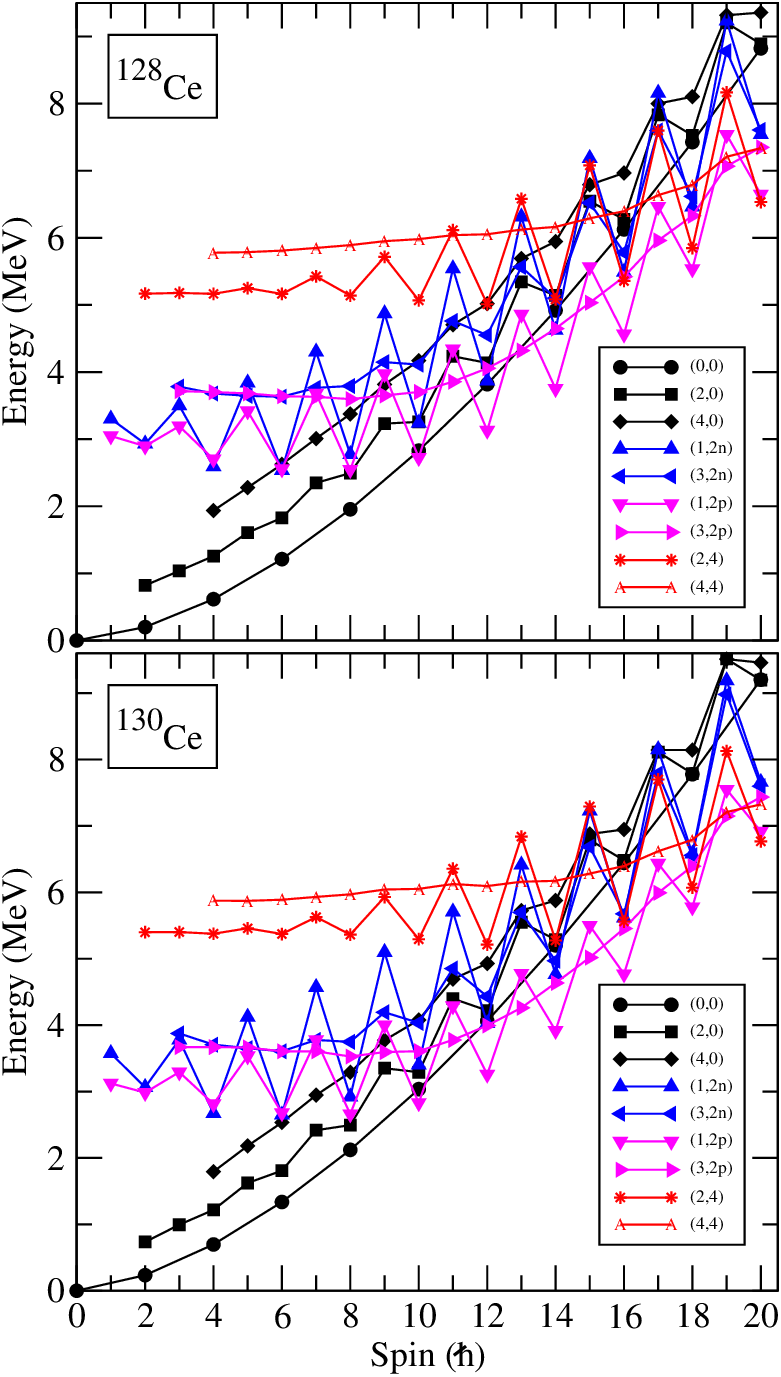}
\includegraphics[totalheight=12cm]{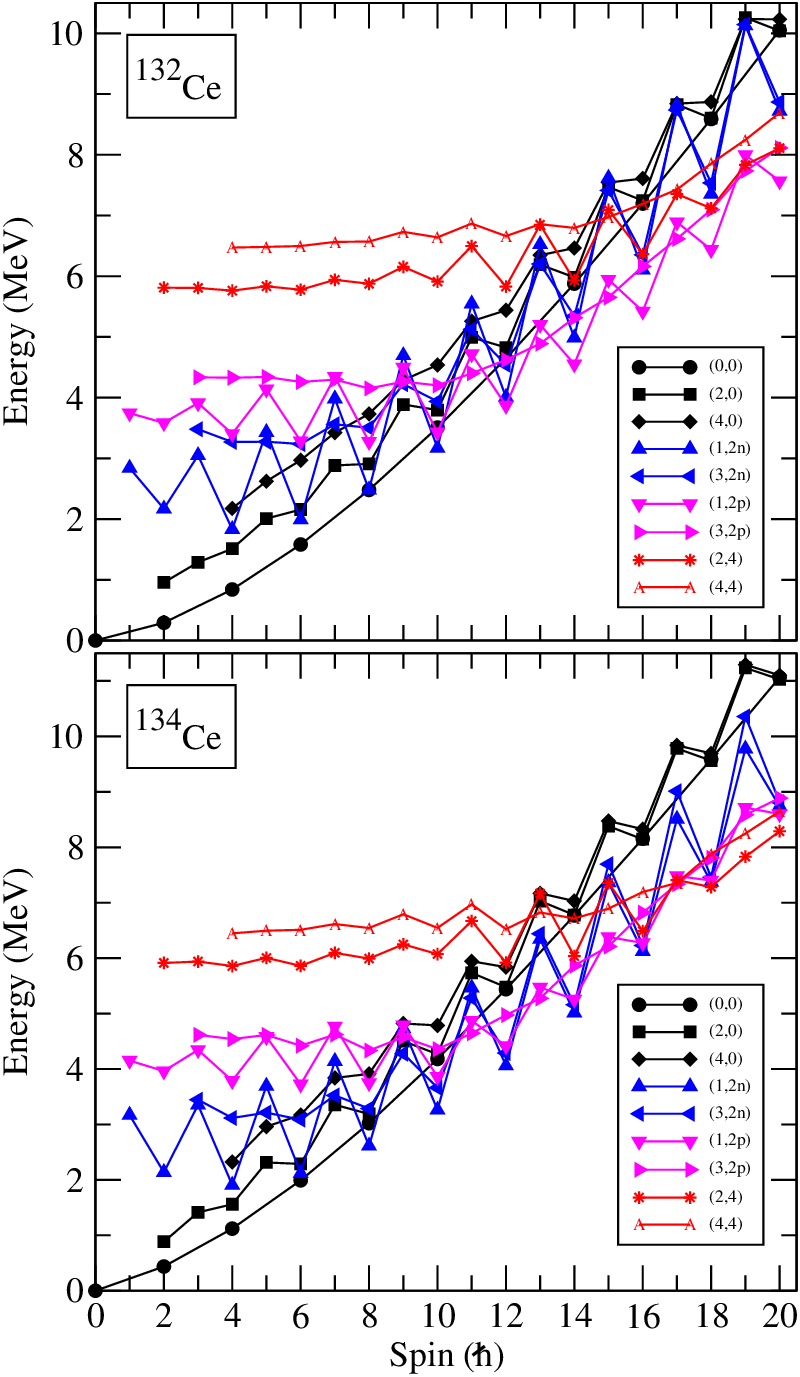}
\caption{(Color online) Band diagrams for $^{128-134}$Ce isotopes.}
\label{fig3}
\end{figure*}

Band diagrams of the studied cerium isotopes from $A=128$ to 134 are
presented in Fig. 3. For $^{128,130}$Ce, the $(2,0)$ bandhead energy
is approximately 0.8 MeV and the $(4,0)$ bandhead is slightly
below 2 MeV. Further, in both nuclei, the $2p$ bandhead is slightly
at a lower excitation energy than the $2n$ bandhead, and due to this
difference, the $2p$ band crosses the ground band earlier than the
$2n$ band. The $2p$ band, $(1,2p)$ crosses the ground band $(0,0)$
in both nuclei at $I= 10$ and the proton structure of the first
crossing is validated by the systematics of the band-crossings
\cite{Rg89}. The g-factor measurement has been done for the $I=10^+$
state of $^{126}$Ce and has a very large value of 1.0, therefore
indicating that the first band-crossing is indeed due to the
alignment of protons for this nucleus. From the known experimental
result and from the systematics of the proton crossing as a function
of mass number, it was concluded in ref. \cite{Rg89} that
$^{128-132}$Ce have proton crossing occurring first.

The band diagrams in Fig. 3 depict $(2,0)$ and $(4,0)$ bandheads for
$^{132,134}$Ce at a similar excitation energy as that of lighter
Ce-isotopes. However, it is noted that the $2n$ bandhead is lower than
the $2p$ bandheads as compared to the $^{128,130}$Ce isotopes. Due
to this lowering of the neutron 2-qp band, the band-crossing
features in the two pictures on the right column in Fig. 3 are
qualitatively different from those of the left column. In
$^{132}$Ce, it is observed that the $2n$ state $(1,2n)$ and $2p$
state $(1,2p)$ become yrast at the same angular momentum of $I=10$
with the neutron band slightly lower than the proton band. However,
for higher angular momenta, the proton band is observed to be favored
in energy, and is yrast up to the highest spin value. For
$^{134}$Ce, the neutron-aligned band $(1,2n)$ crosses the ground
band $(0,0)$ at $I=8$, and this band is yrast up to $I=16$. Above
this spin value, it is noted that the 4-qp band $(2,4)$ with both
protons and neutrons aligned becomes favored.

\begin{figure*}[htb]
\includegraphics[totalheight=12cm]{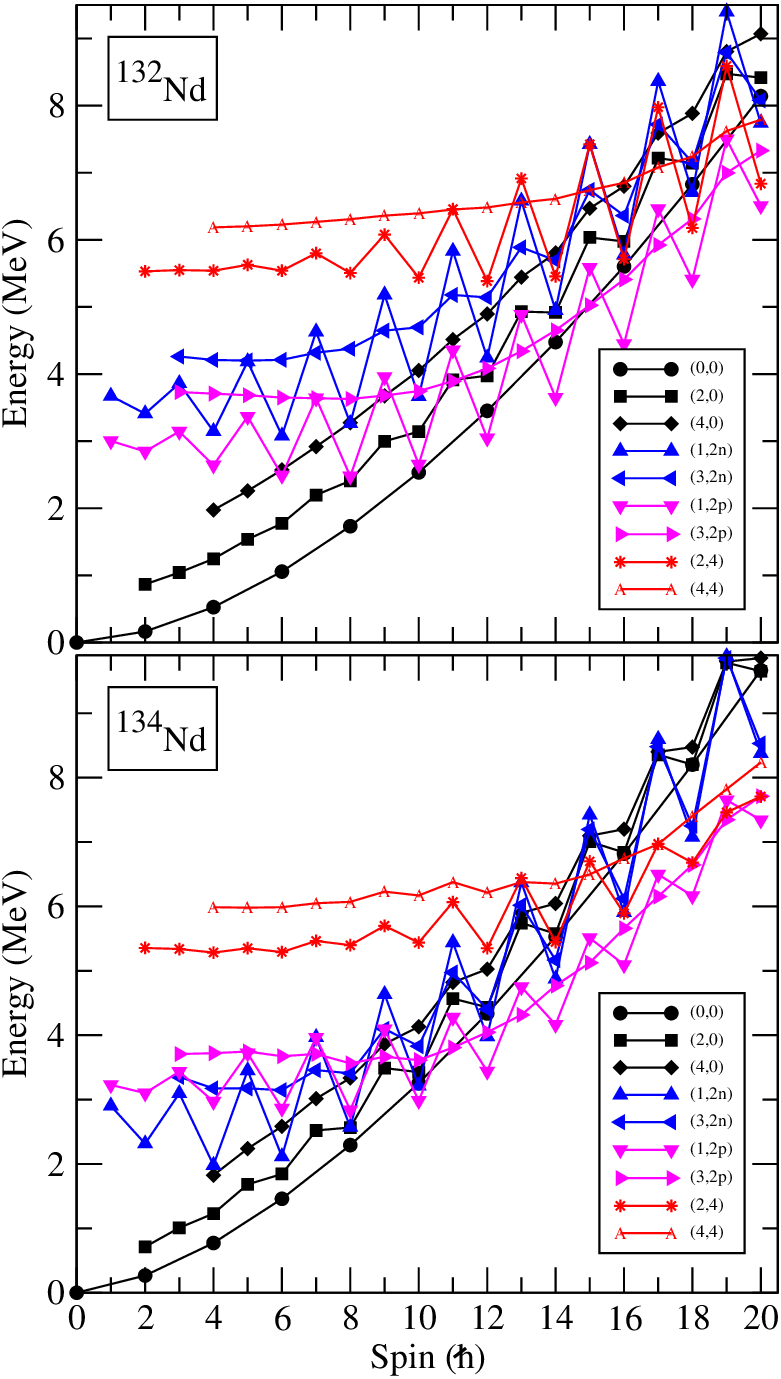}
\includegraphics[totalheight=12cm]{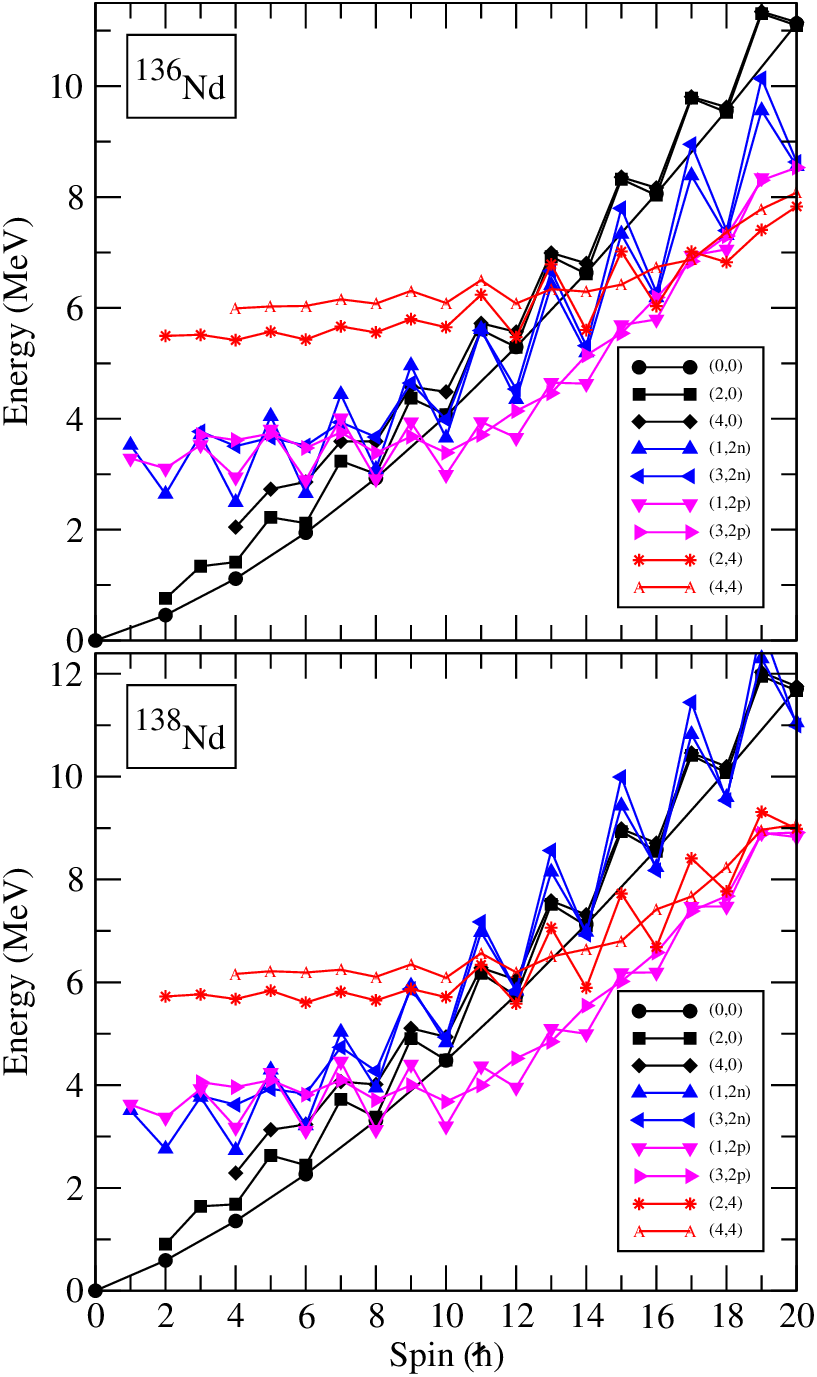}
\caption{(Color online) Band diagrams for $^{132-138}$Nd isotopes.}
\label{fig4}
\end{figure*}

Band diagrams of the Nd isotopes are shown in Fig. 4. It is seen
that the $(2,0)$ and $(4,0)$ bandheads are, respectively, at 1 and 2
MeV excitation energy in $^{132}$Nd. The $2p$ aligned bandhead is
lower in energy than the $2n$ band with the result that the proton
band $(1,2p)$ crosses the ground band $(0,0)$ at $I=12$, and becomes
yrast for all the spin values above it. For $^{134}$Nd, the $2n$
bandhead is now lower in energy than the $2p$ one, and at $I=10$
both these aligned bands cross the ground band and it is, therefore
expected that this nucleus should depict forking of the ground state
band into two s-bands. Above the band-crossing region in $^{134}$Nd,
the yrast even-$I$ states originate from the $(1,2p)$ and the
odd-$I$ states arise from the $(3,2p)$ band. The lowest two bands
above $I=12$ originate from the same $2p$ states and, therefore,
will have positive g-factors, opposite to that of the $^{134}$Ce
case.

For $^{136,138}$Nd in Fig. 4, the $\gamma$-bands are very close to
the yrast line. In $^{136}$Nd, the band-crossing occurs at $I=8$,
and for $^{138}$Nd it is at $I=10$. In both nuclei, proton- and
neutron-aligned bands cross the ground band and, therefore, it is
expected that both nuclei should show forking of the ground state
band into two s-bands. Further, it is noted that in $^{138}$Nd the
$(3,2p)$ band is also very low in energy and is the first excited
band above $I=10$ for even-spin states. For odd-spin states, it
forms the yrast band.

In the mass region under investigation, there have been observations
of high-spin states at low excitation energies. We previously
mentioned $^{134}$Ce as a particular example where two lowest $10^+$
states were detected at excitation energy 3.2086 MeV and 3.7193 MeV,
respectively. In Fig. 5, we collect all the experimentally known
$10^+$ states for nuclei studied in this paper. At first glance,
these could be bandheads of some multi-qp bands associated with
high-$K$ configurations. However, a close look indicates that there
are no such high-$K$ configurations available around the Fermi
surfaces. From our band diagrams shown in Figs. 3 and 4, we see that
in nuclei with $N=76$ and 78, more than one band crosses the ground
band at $I=10$. The crossing bands are not high-$K$ bands, but bands
having low-$K$ configurations.

\begin{figure}[t!]
\includegraphics[totalheight=14cm]{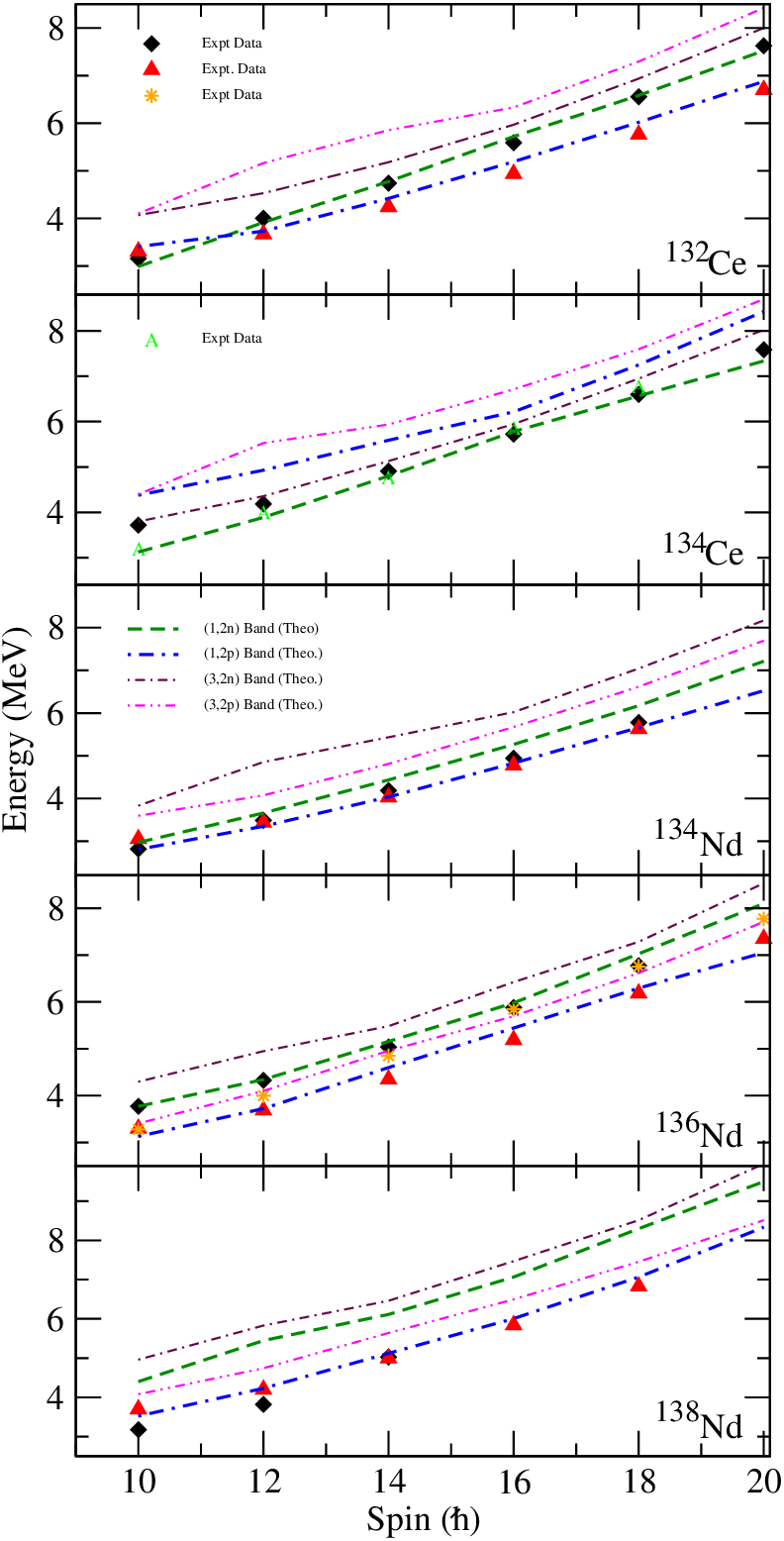}
\caption{(Color online) Four theoretical bands with the main
component from $(1,2n)$, $(1,2p)$, $(3,2n)$, and $(3,2p)$,
respectively. Only the spin range from $I=10$ to 20 is shown in
which these bands are low in energy. Available data in
$^{132,134}$Ce and $^{134,136,138}$Nd are compared with the
calculation.} \label{fig5}
\end{figure}

The two $I=10^+$ states in $^{134}$Ce were long ago identified
experimentally. However, the structure of these $10^+$ states
remained a puzzle. The magnetic moment of both states has been
measured, and it has been found that both have negative g-factors
\cite{Ze82}. This suggests that both the $10^+$ states have a
neutron structure. This was a surprising finding because normally
two lowest-lying 2-qp states should separately belong to
2-quasineutron and 2-quasiproton states. In ref. \cite{Rg89}, an
explanation was proposed. Now looking at our Fig. 3, one can easily
see that in $^{134}$Ce, apart from the 2$n$ band $(1,2n)$ and the
2$p$ band $(1,2p)$, the $\gamma$-band built on the $2n$ band
$(3,2n)$ also crosses the ground band $(0,0)$ at $I=10$. Thus at
$I=10$, there are at least three states below that of the ground
band, which are (from lower to higher energies) $(1,2n)$, $(3,2n)$,
and $(1,2p)$. It is very interesting that for this nucleus, the
neutron 2-qp band based on $\gamma$-vibration $(3,2n)$ is predicted
to appear lower in energy than the proton 2-qp band $(1,2p)$.
Therefore, the lowest two $I=10^+$ states, one being a 2-qp state
built on the ground state and the other a 2-qp state on the
$\gamma$-vibration, have both neutron configuration. This is
consistent with the g-factor measurement \cite{Ze82}, and thus
naturally explains the structure of the two $10^+$ bands.

\begin{figure}[t!]
\includegraphics[totalheight=10cm]{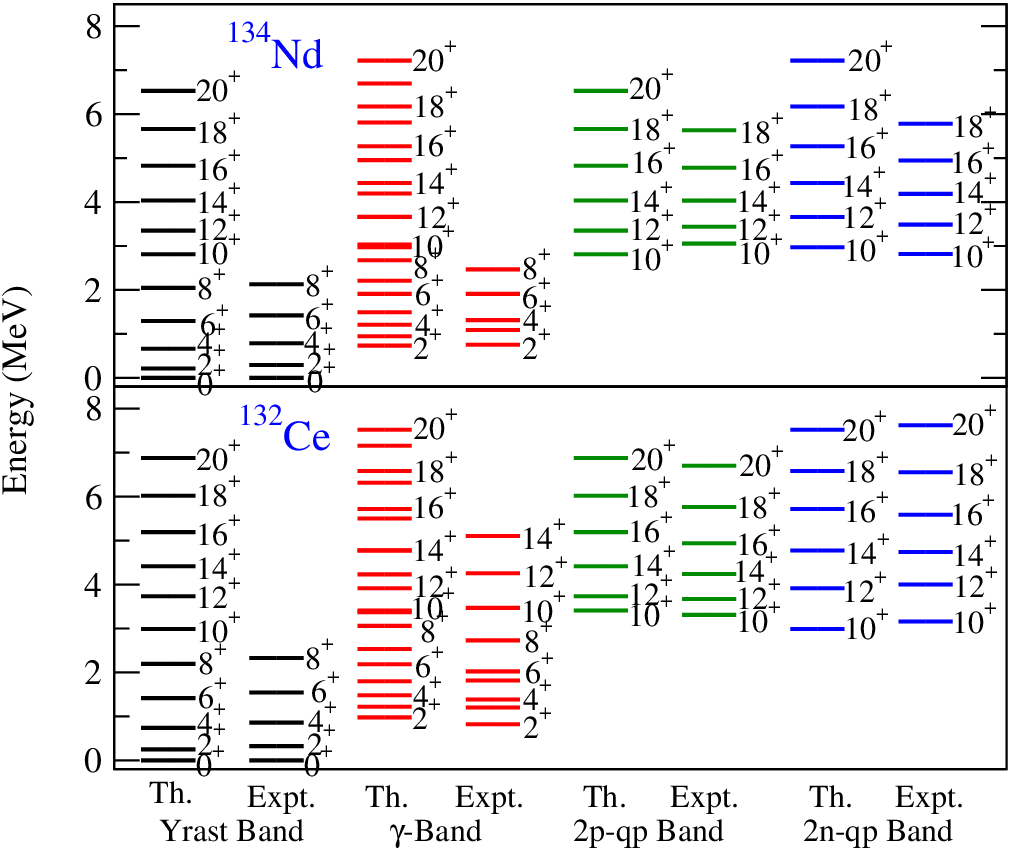}
\caption{(Color online) Level schemes for $^{132}$Ce and $^{134}$Nd.
Comparison between the calculated levels and experimental data is
made for four bands: the yrast bands, the $\gamma$ bands, and the
$I=10^+$ 2-qp bands based on $\gamma$ vibration.} \label{fig6}
\end{figure}

Fig. 5 summarizes the results regarding the theoretical $\gamma$-
bands built on 2-qp states obtained after diagonalization of the
shell-model Hamiltonian. The calculated bands associated with the
four most relevant and dominant states are displayed: (1) $(1,2n)$,
the 2-qp neutron $K=1$ state built on the qp vacuum state; (2)
$(1,2p)$, the 2-qp proton $K=1$ state on the vacuum state; (3)
$(3,2n)$, the 2-qp neutron $K=3$ state on the collective
$\gamma$-vibration; and (4) $(3,2p)$, the 2-qp proton $K=3$ state on
the $\gamma$-vibration. Only even spin members (energetically
favored) of each band are shown and compared with available data. It
is quite interesting to note that the relative position of the four
bands varies in each nucleus. We have discussed the case of
$^{134}$Ce where the lowest two are both neutron states. From Fig.
5, our calculation further predicts that in $^{136,138}$Nd, one may
observe two lowest $I=10^+$ states but with a proton structure. The
g-factor measurement in these two nuclei is expected to lead to
large, positive values, in sharp contrast to those in $^{134}$Ce.

Finally, in order to depict the comparison between the experimental
and theoretical energies more clearly, we plot in Fig. 6 the
calculated and experimental level energies for the yrast bands, the
$\gamma$ bands, and the $I=10^+$ 2-qp bands based on $\gamma$
vibration. We have chosen two nuclei, $^{132}$Ce and $^{134}$Nd, as
examples and the results are similar for other studied nuclei. It is
quite evident from this figure that not only the levels within each
band, but also the relative positions of the bands, are reasonably
well reproduced by the TPSM approach.

\section{Summary}

The results presented in this work suggest that multi-qp states in a
triaxially deformed well can exhibit much more fruitful band
structures. This is because with triaxiality, a single configuration
contains a rich mixture of many possible $K$-states and after
angular momentum projection, each one of them corresponds to a
rotational band. It was pointed out in ref. \cite{YS02} that the
projected triaxial vacuum state alone can already produce the
collective ground state band, $\gamma$-band, 2$\gamma$-band, etc.
Similarly, a projected 2-qp state can give rise to 2-qp bands with
$K=1,3,5,\cdots$. If the $K=1$ band is a 2-qp band based on the
ground state, then the $K=3$ band can be understood as a 2-qp band
based on the $\gamma$- vibration, and the $K=5$ band as a 2-qp band
based on the 2$\gamma$ state, etc. This pattern can be clearly seen
when multi-qp configurations constructed from a triaxially deformed
well are projected on good angular momentum. The picture thus
extends the simple surface $\gamma$ oscillation in deformed nuclei
where paired nucleons vibrate coherently in the deformed vacuum
state.

Summarizing the present work, multi-qp band structures in some
neutron-deficient Ce and Nd isotopes has been studied using the
extended triaxial projected shell-model approach. It has been
demonstrated that $\gamma$-band built on the 2-qp configurations can
modify the band-crossing features in these nuclei. The 2-qp
$\gamma$-band with $K=3$ are shown to be energetically favored for
some angular-momentum states and form the first excited bands in
nuclei studied in the present work. For $^{134}$Ce, it is shown that
the lowest two $I=10^+$ states originate from the same
2-quasineutron configuration and sheds new light on the observation
of negative g-factors for the two states. Further, it is predicted
that the lowest two $I=10^+$ states in $^{136,138}$Nd originate from the
same $2p$ configuration and both these states should have positive
g-factors. The present results have enriched the concept of
$\gamma$-vibration and we hope that, in the future, more states of such
kind will be identified experimentally.

\section{Acknowlegments}
Research at ORNL is supported by the Division of Nuclear Physics,
U.S. Department of Energy, under Contract DE-AC05-00OR22725 with
UT-Battelle, LLC. Research at SJTU is supported by the National
Natural Science Foundation of China under contract No. 10875077 and
by the Chinese Major State Basic Research Development Program
through grant 2007CB815005.

\end{document}